# Giant superchiral field at merged bound states in the continuum

Hamdi Barkaoui[1], Kang Du[1], Yimu Chen[1], Shumin Xiao[1,2,3], Qinghai Song[1,2,3,*]

1. Ministry of Industry and Information Technology Key Lab of Micro-Nano Optoelectronic Information System, Harbin Institute of Technology; Shenzhen 518055, P. R. China.
2. Pengcheng Laboratory; Shenzhen, 518055, P. R. China.
3. Collaborative Innovation Center of Extreme Optics, Shanxi University; Taiyuan, Shanxi 030006, P. R. China

*Corresponding author. Email: qinghai.song@hit.edu.cn;

**Abstract:** Superchiral field is highly important for ultrasensitive detection of chiral objects and has been intensively explored. Despite the progress, the construction of electromagnetic field in achiral nanostructures with giant chirality ($C$) is fundamentally restricted. Here, we demonstrate a novel approach to construct superchiral field in achiral dielectric metasurfaces. Due to the quantum spin-hall effect, each symmetry-protected bound state in the continuum (BIC) exhibits the ability to support a superchiral field in vicinity of Γ point. By merging two BICs with orthogonal polarizations, all the criteria for a superchiral field are satisfied and the dramatic enhancement of chirality becomes more robust. Using typical photonic crystal slabs, we have produced a superchiral field with enhancement factor ($C/C_{\text{CPL}}$) orders of magnitude higher than state-of-the-art in nanostructures. By further introducing active chiral medium, a chiral mode splitting associated with giant enhancement of circular dichroism (CD) has been formed by the chiral light-matter interaction.

Chirality is an inherent and ubiquitous property of biological systems as well as chemically synthesized drugs [1]. While their transition frequencies are degenerate, enantiomers often exhibit strongly opposing behaviors in their metabolic uptake, and thus the identification of chiral molecules become critical for chemistry, biology and pharmacology [2]. By exploiting the interaction between chiral molecules and chiral electromagnetic field, circular dichroism (CD) spectroscopy has emerged as a versatile technique for the detection of chiral species [3]. Due to the weak CD signal of chiral molecules, the initial spectroscopic experiments are only applicable to very high concentrations of analytes. In the early 2010s, Tang and Cohen have proposed and experimentally demonstrated a new type of field with chiral density much higher than circular polarized light (CPL), known as superchiral field now [4,5]. The enhanced interaction between superchiral field and matter can greatly alleviate the dependence on the density of chiral compounds and has triggered a new research direction in nanophotonics.

In principle, the optical chirality of electromagnetic field is given by $C = -\frac{1}{2}\varepsilon_0 Im(E^* \cdot B)$ [6], where E and B are the complex time varying electric and magnetic field vectors. The components of electric and magnetic fields should fulfill three conditions to create a giant superchiral field, i.e., i) strong and spectral overlap, ii) parallel and spatial overlap, and iii) $\pi/2$ out of phase [7]. Over the past decade, many efforts have been made to increase chiral density using engineered fields in artificial nanostructures [8–14]. Most initial schemes, however, are irrelevant to the identification of chiral molecules due to their high optical loss [15–18] and extrinsic CD generated by the nanostructures [19]. In 2019, E. Mohammadi et al., have proposed superchiral field in dielectric achiral nanostructures using spatially and spectrally overlapped magnetic-dipole and electric-dipole resonances [20]. Soon after, under the illumination of two beams with a $\pi/2$ phase shift, T. Wu et al., further increased the enhancement factor $C/C_{CPL}$ to ~ 250 at an exceptional point [21]. Despite the progress on spatial and spectral overlap, little attention has been paid to the quality (Q) factor that is critical for the construction of ***strong*** electromagnetic fields.

Bound state in the continuum (BIC) is a non-radiating solution of a wave equation [22] and has been extensively studied in diverse systems [23–25]. The introduction of BICs to photonics is revolutionizing the design of optical cavities and their internal light–matter interactions [26–30]. A variety of unique functions have been demonstrated at the BICs, e.g., hyperspectral imaging [31], low threshold microlasers [32], ultrafast control [33,34], and efficient harmonic generation [35,36]. In 2019, it is recognized that the quantum spin-Hall effect (QSHE) can effectively excite the high Q quasi-BIC modes and produce directional propagating fields in the photonic crystal slabs [37]. Herein, we employ such unique characteristic of BICs to produce superchiral field. By merging two BICs with orthogonal polarizations [29,38], we further show all the requirements for a giant superchiral field have been matched simultaneously. The dramatic enhancement of $C/C_{CPL}$ becomes more robust and an interesting chiral mode splitting emerges.

The schematic of our metasurface is illustrated in Fig.1(a). It is composed of square-latticed air holes in $Si_3N_4$ membrane on a glass substrate. The lattice size and hole radius are $a_0$ and $R$, respectively. The thickness of slab is fixed at 180 nm. The band structure of the metasurface has been calculated and plotted in Fig. 1(b). A series of TE-like and TM-like resonances can be observed around Γ point [39]. Only four of them have dramatically suppressed radiation loss, corresponding to symmetry protected BIC modes. Here and below, we will focus on $TE_3$ and $TM_2$ bands, which are referred as TE and TM bands.

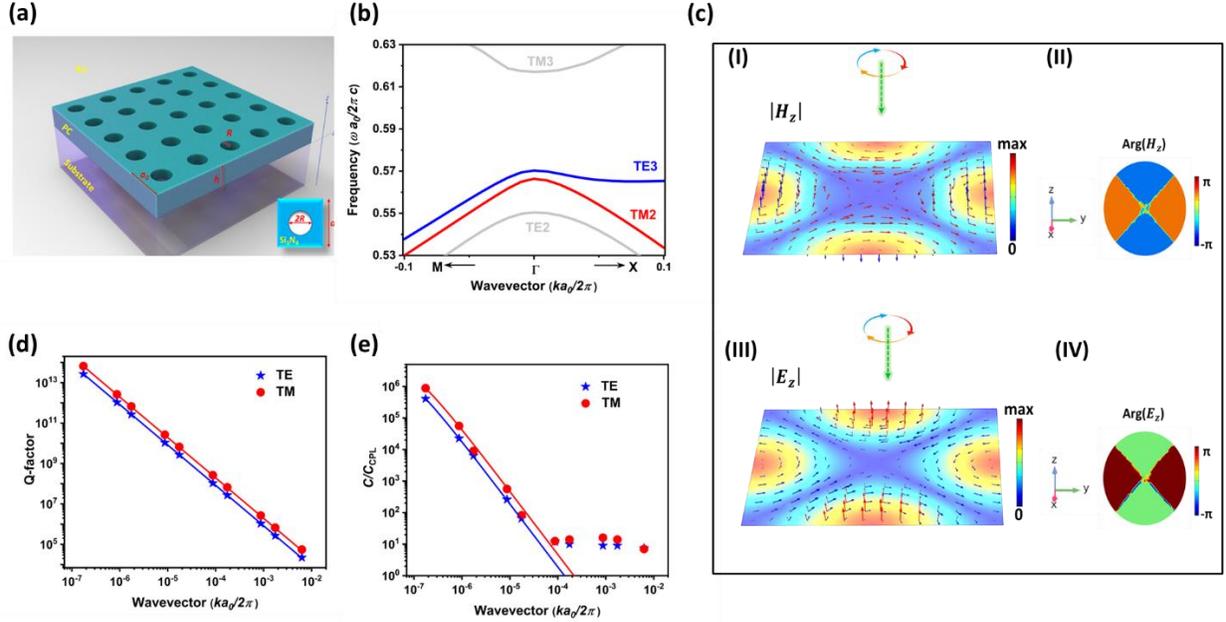

**Figure1:** (a) Sketch of the designed photonic crystal. The structural parameters are $a_0 = 321\ nm$, $R = 81\ nm$, and $h = 180\ nm$, where $a_0$, $R$ and $h$ design the unit cell period, the net radius and the slab thickness, respectively. (b) Corresponding photonic band structure of the photonic crystal in the vicinity of Γ point. The modes of interest are colored in blue (TE) and red (TM). No leaky modes are represented here. (c) Excitation with a CPL source. Amplitude (I) and phase (II) of the longitudinal component of the electric field for the studied TE mode. Amplitude (III) and phase (IV) of the longitudinal component of the magnetic field for the studied TM mode. Blue and red arrows correspond to the magnetic and electric fields, respectively. (d) Wavevector dependence of the Q-factor near Γ point for the TE (blue stars) and TM (red circles) modes. The solid lines represent $1/k^2$ fitting curves. (e) Wavevector dependence of the chiroptical parameter enhancement near Γ point for the TE (blue stars) and TM (red circles) modes. The solid lines represent $1/k^2$ fitting curves.

In principle, the BICs are decoupled from the external continuum and hard to be excited [40]. With the assistance of QSHE, the incident CPL can be partially coupled into the BIC or quasi-BIC modes (see panels I and III in Fig. 1(c)). Although the coupling efficiency is limited at the BICs, their ultrahigh Q factors are able to trap and accumulate electromagnetic field for an extremely long time ($\tau \propto \frac{Q}{\omega}$), resulting in extremely large E or H [41]. Figure 1(d) shows the wavevector dependent Q factors of the two BICs. Both TE and TM BICs are quadratically dependent on the incident angle (see fitted lines in Fig. 1(d)) and above $10^{13}$ in the vicinity of Γ point [42]. Due to the finite thickness, the BICs in three-dimensional nanostructures are TE-like and TM-like modes. Consequently, there must be a tiny associated H and E components parallel to their dominant E and H fields. According to the definition of chirality $C$, it is easy to know that the chirality of electromagnetic field inside the photonic crystal slab should be enhanced. And the corresponding enhancement factor increases dramatically approaching the Γ point. Figure 1(e) summarize the numerically calculated enhancement factor $C/C_{CPL}$ as a function of the wavevector $k$. It is apparent that the enhancement factor increases exponentially approaching the Γ point. The $C/C_{CPL}$ at $ka_0 = 1.7 \times 10^{-7}\pi$ is even as high as $10^6$, more than 3 orders of magnitude larger than previous reports.

The enhancement of $C$ is the vicinity of Γ point is primarily caused by the high Q factor, which is inversely proportional to the square of perturbations. Apparently, the enhancement $C/C_{CPL}$ follows the same proportionality rule as the Q factor for small $k$ (see fitted line in Fig. 1(e)). Genuinely, for the TE mode for example, both components $H_z$ and $E_z$ are strong and well-defined around the center of the momentum space, leading to an almost constant phase shift between these factors (see Supplementary Information [43]). However, when the deviation from Γ point becomes more important, only $H_z$ component remains strong and well-defined. Consequently, the phase shift will experience some fluctuations and will play a more important role in the value of the enhancement of $C$. As depicted in Fig. 1(e), the enhancement factor reduces rapidly to only around 10 at $ka_0 = 6.3 \times 10^{-3}\pi$, corresponding to an incident angle of 0.36 degrees. Such a giant superchiral field ($C/C_{CPL} > 1000$) is hard to be achieved. As a result, new mechanism must be introduced to improve the robustness.

The magnetic (H) and electric (E) field distributions of the TE and TM BICs are very similar and parallel to each other (panels I and III in Fig. 1(c)). Meanwhile, their phase differences are also fixed π/2 as well (panels II and IV in Fig. 1(c)). In this sense, if these high Q modes approach one another, they are able to fulfill all the requirements for a giant chirality. Consequently, the chirality of quasi-BIC can be further enhanced and it is possible to construct giant superchiral field at relatively large angle $k$.

To verify this possibility, we have merged the TE and TM BICs numerically. Owing to their different distributions of evanescent fields, the frequency difference between two BICs can be modified by either changing the hole size, lattice size, or the thickness of membrane. Here we vary both $a_0$ and $R$ in the following way: $a_0 = p + l/1.7$ and $R = r + l$, with $p = 320$ nm, $r = 78$nm, and a variable parameter $l$. Figure 2(a) summarizes the normalized frequencies of two BICs as a function of $l$ at an incident angle deviated from Γ point to $\theta = 0.36°$. With the increase of $l$, we can see that two modes gradually approach one another and cross at $l = 0$ nm. Meanwhile, the corresponding chirality $C$ of two modes have also been calculated by taking their electric field (E) and magnetic field (H) into accounted. The enhancement factor ($C/C_{CPL}$) are plotted in Fig. 2(b).

Due to their restricted Q factors, the individual quasi-BICs only have $C/C_{CPL} \sim 10$. Once two BICs merge, the enhancement factor has been dramatically enhanced by almost two orders of magnitude. All these numerical results are consistent with the above analysis very well.

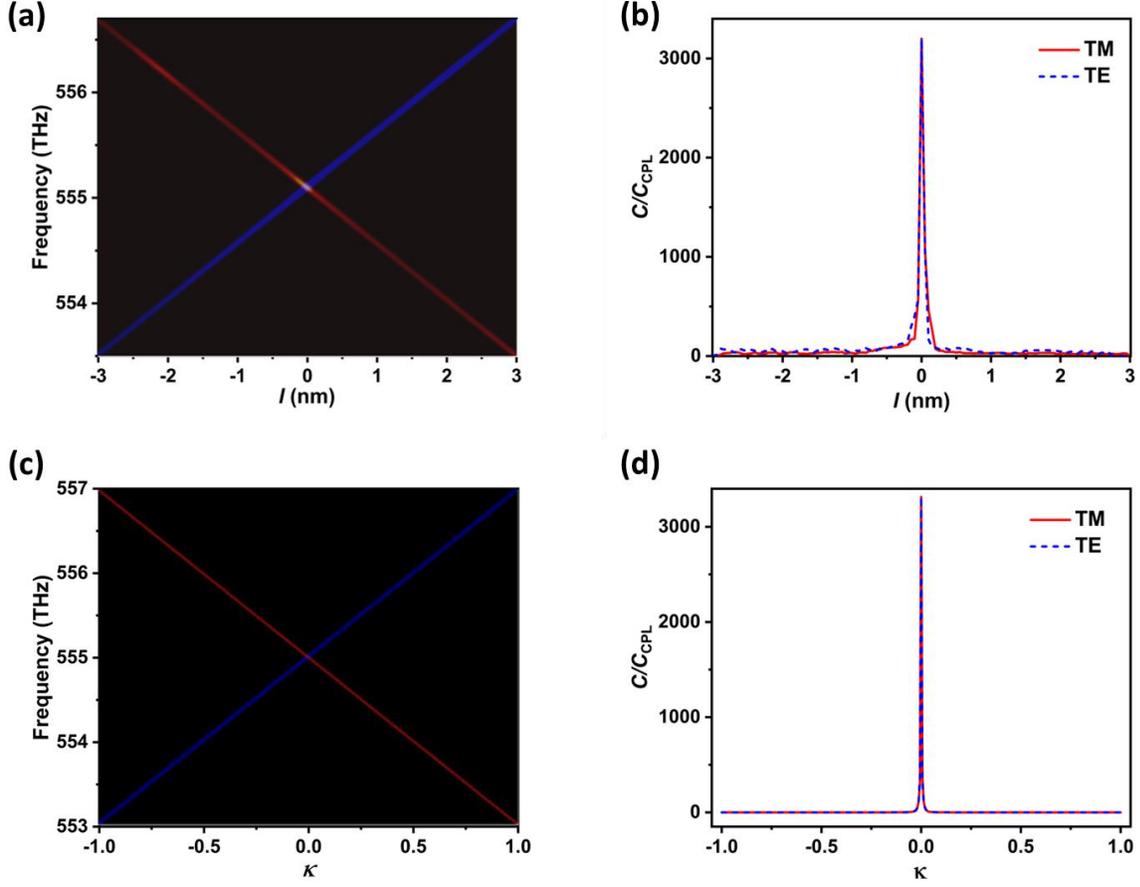

**Figure2:** (a) Numerical simulations of the positions of TE and TM quasi BICs by tuning the parameter l. The merging point occurs at a frequency of 555 THz for $l = 0$. (b) The corresponding $C$ enhancement values. $C$ increases dramatically at $l = 0$. (c) Analytical calculations of the positions of a two coupled quasi BICs by tuning the coupling constant κ. The merging point was set to be the same as the numerical one. (d) The corresponding C enhancement values. $C$ increases dramatically at $\kappa = 0$. The parameters used here are $\gamma_E = 4 \times 10^{-3}$ THz $\gamma_E = 3 \times 10^{-3}$ THz, $W_{EM} = 555$ THz, $\Delta W_{EM} = 10^{-2}$ THz and $\sin(\phi) C_{CPL} = 8$ S.I.

For quantitatively understanding such enhancement, we have theoretically analyzed the system with temporal coupled mode theory (TCMT) [44]. We consider one TE and one TM modes with amplitudes $A_E$ and $A_M$. Their frequencies are expressed as $\omega_E + i\gamma_E$ and $\omega_M + i\gamma_M$, respectively. Assuming $S_+$ and $S_-$ as incoming and outgoing waves [45], the amplitudes of two modes can be achieved by the TCMT with the following equations,

$$\frac{dA}{dt} = (i\Omega - \Lambda)A + D^T|s_+\rangle, \tag{1}$$

$$|s_-\rangle = C|s_+\rangle + DA. \qquad (2)$$

Here $A = (A_E \quad A_M)^T$, $\Omega = \begin{pmatrix} \omega_E & \kappa \\ \kappa & \omega_M \end{pmatrix}$, $\Lambda = \begin{pmatrix} \gamma_E & \beta \\ \beta & \gamma_M \end{pmatrix}$, $D = \begin{pmatrix} d_{E1} & d_{M1} \\ d_{E2} & d_{M2} \end{pmatrix}$ and $C = \begin{pmatrix} r_d & t_d \\ t_d & r_d \end{pmatrix}$. The near-field and far field coupling coefficients are defined as $\kappa$ and $\beta$, respectively. The elements $d_{Ei}$ and $d_{Mi}$ in D are the radiation of TE and TM modes to the $i$-th port. $r_d$ and $t_d$ represent the direct reflection and transmission coefficients. The Hamiltonian H of the system is $H = \Omega + i\Lambda$. Then the eigenvalues can be expressed as

$$E_\pm = W_{EM} + i\Gamma_{EM} \pm \sqrt{(\Delta W_{EM} + i\Delta\Gamma_{EM})^2 + 4(\kappa + i\beta)^2}, \qquad (3)$$

where $W_{EM} = {\omega_E + \omega_M}/{2}$, $\Gamma_{EM} = {\gamma_E + \gamma_M}/{2}$, $\Delta W_{EM} = {\omega_E - \omega_M}/{2}$ and $\Delta\Gamma_{EM} = {\gamma_E - \gamma_M}/{2}$.

Considering the incident light from port-1 along -$z$ direction, Eq. (1) can be rewritten as

$$\begin{pmatrix} A_E \\ A_M \end{pmatrix} = [-i(-\omega I + \Omega) + \Lambda]^{-1} D^T \begin{pmatrix} s_{1+} \\ 0 \end{pmatrix}. \qquad (4).$$

From the definition of chiroptical parameter $C$, it is proportional to the amplitudes of the two components. Consequently, the chirality can be given by

$$C = -\frac{1}{2}\varepsilon_0 Im(E^* \cdot B) = -\frac{1}{2}\varepsilon_0 A_E A_M \sin(\phi) s_{1+}^2$$
$$\approx \frac{4\sqrt{\gamma_E \gamma_M}(\gamma_E + i\omega_E - i\omega)(\gamma_M + i\omega_M - i\omega)}{[\kappa^2 - (\gamma_E - i\omega_E + i\omega)(\gamma_M - i\omega_M + i\omega)]^2} \sin(\phi) C_{CPL}. \qquad (5)$$

Here $C_{CPL}$ is the chirality brought by the light source and $\phi$ is the phase difference between the TE and TM modes. The detailed derivations can be seen in the supplementary materials [43]. By taking the parameters of the calculated eigenfrequencies, the chirality $C$ has been calculated with Eq. (5) and plotted in Figs. 2(c) and 2(d). With the increase of the coupling coefficient $\kappa$, we are also able to see the crossing of two quasi-BICs. At the merging point, the chirality $C$ has been dramatically enhanced from ~ 30 to more than 3000, consistent with the numerical simulation very well.

One very intriguing information from Eq. (5) is that the enhancement factor $C/C_{CPL}$ can go to infinity when the imaginary frequencies $\gamma_{E,M}$ gradually decrease, i.e., the merging point gradually approach Γ point. For that purpose, we tried to reduce $k_M$, the radial wavevector coordinate describing the circle of merged points in the 2D Brillouin zone as shown in Figure 3(a), by slightly shifting the net radius. Actually, thanks to the $C_4$ symmetry of the system, both TE and TM bands are forming a parabolic shape, hence the iso-frequency contours will be circularly distributed in the $k$ space [46]. In the case where $a_0 = 320$ nm and $R = 78$ nm, we obtain $k_M a_0 = 63.10^{-4}\pi$. The merging point changes with the structural parameters as well and can also approach to the Γ point. Figure 3(b) shows the dependence of $C/C_{CPL}$ on the position of merging point ($k_M$). We can see that the enhancement factor also increases dramatically when $k_M \to 0$. For instance, at $k_M a_0 = 17.10^{-4}\pi$, corresponding to an elevation angle of incidence of 0.1 degrees, the chiroptical parameter reaches an enhancement of about 57000. Besides, at every value of $k_M$, the enhancement at the merged point is obviously much higher than those of the isolated BICs (squares in Fig. 3(b)), proving the prevailing of this particular point. Moreover, the $C/C_{CPL}$ curve is inversely proportional to the quartet of $k_M$ (see solid line in Fig. 3(b)), as it overlaps both enhancement of the independent modes, which follow a $1/k^2$ as we stated previously. By contrast to the isolated BIC, the phase shift of the merged BIC is constant and the enhancement curve can be well fitted even for relatively large angles, since both the components of the electromagnetic field are well defined.

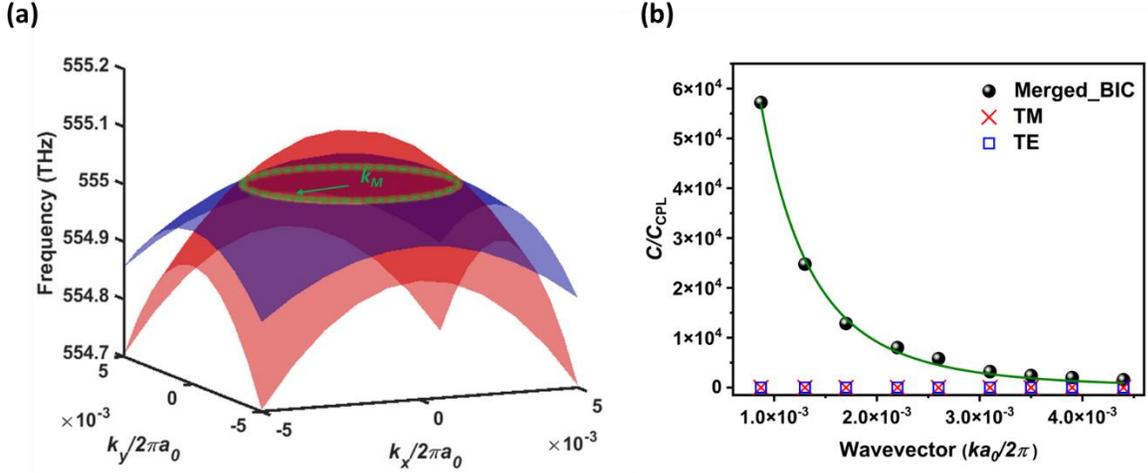

**Figure 3:** (a) Illustration of the TE (blue) and TM (red) bands the vicinity of Γ point in the 2-dimensional momentum space for $l = 0$. The green dashed line displays the circle of wavevectors where the merging of the two modes occurs. (b) Comparison of wavevector dependences of $C/C_{CPL}$ between the merged (dots) and isolated (cross and squares) BICs. The green line represents the 1/$k^4$ fitting curve.

The superchiral field in achiral nanostructure is particularly important for CD measurement. When chiral molecules are considered, their optical response can be described by the following constitutive relationships,

$$\begin{pmatrix} D \\ B \end{pmatrix} = \begin{pmatrix} \epsilon & -i\chi/c \\ i\chi/c & \mu \end{pmatrix} \begin{pmatrix} E \\ H \end{pmatrix} \quad (6)$$

$$\text{and } \nabla \times \begin{pmatrix} E \\ H \end{pmatrix} = \frac{\partial}{\partial t} \begin{pmatrix} -B \\ D \end{pmatrix}, \quad (7)$$

where is $\chi$ the Pasteur parameter (purely real) and $c$ is the speed of light in the vacuum, $\epsilon$ is the permittivity and $\mu$ is the permeability [47]. For conventional chiral absorption, the dramatically enhanced $C/C_{CPL}$ at merged BIC can improve the amplitude CD response. When the active chiral media is considered, the situation becomes more interesting. Figure 4(a) displays the numerically calculated transmission coefficients spectra for different types of molecules (R-, S- and racemic) under circular polarized light source. Here the material parameters are $n = 2.04$ and $\chi = 10^{-4}$, the incident angle is fixed 0.36°. The transmission spectrum for the R-enantiomer under RCP excitation for instance (blue curve in Fig. 4a), the transmission spectra at the merged BIC exhibits a frequency mode splitting comparing to the achiral material [48]. This mode splitting is perfectly symmetric regarding the achiral entity transmission dip $\omega_0$ (black curve in Fig. 4a) and its magnitude is around $10^{-2}$ THz. However, the transmission amplitudes for both dips are remarkably disparate with a much lower transmission in the $\omega < \omega_0$ region. The situation is symmetric under LCP excitation (red curve in Fig. 4a) where the transmission is much lower in the $\omega > \omega_0$ region. Hence, for one spectral region, the differential transmission, defined as $\Delta T_{PC} = t_{++}^2 - t_{--}^2$, is more than three orders of magnitude larger (red curve in Fig. 4(b)). These observations hold true for the L-enantiomer but with opposite sign of the CD comparing to the R-enantiomer.

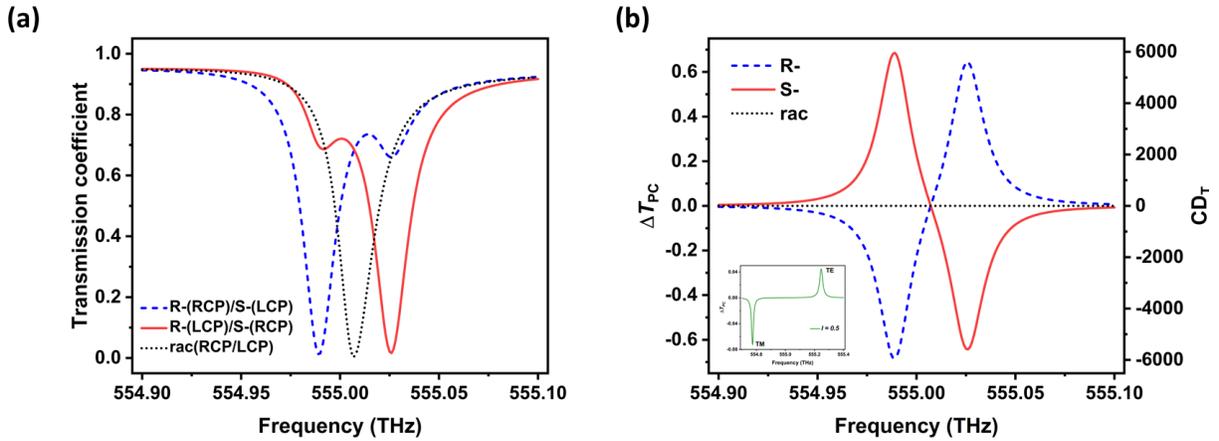

Figure 4: a) Transmission coefficient spectra of a photonic crystal made from R-enantiomer, S-enantiomer and racemic molecules under RCP/LCP source excitations for $l = 0$. b) Differential transmission of the chiral molecules between RCP and LCP excitations and the corresponding circular dichroism enhancement for $l = 0$. The inset shows the differential transmission for $l = 0.5$, where the BICs are separated.

The chiral mode splitting originates from the interaction of the chiral medium with the constructed superchiral field. According to Eq. (6), it is easy to know that the frequency splitting $\Delta\omega$ is closely related to both of the Pasteur parameter and the local electromagnetic field. At the merged BIC, both **E** and **H** are enhanced simultaneously and thus the mode splitting becomes quite apparent. Started with the TCMT, the differential transmission has been theoretically fitted. It is obvious to see that both the wavelengths and the amplitudes can be well matched. For a direct comparison, the chiral mode splitting for an individual BIC has also been studied with the same Pasteur parameter. As depicted in SI [43], the mode splitting cannot be distinguished from the transmission spectrum and the amplitude of differential transmission is almost one order of magnitude smaller (inset of Fig. 4b).

In summary, we have demonstrated a new mechanism to construct superchiral field in achiral dielectric metasurfaces. By merging two BICs with orthogonal polarizations, all the criteria for a superchiral field can be satisfied, i.e., i) strong and spectral overlap, ii) parallel and spatial overlap, and iii) $\pi/2$ out of phase. This discovery breaks the conventional belief of superchiral generation in the realm of chiral photonics. Using typical photonic crystal slabs, we have produced a superchiral field with enhancement factor ($C/C_{\text{CPL}}$) orders of magnitude higher than state-of-the-art in nanostructures. The enhancement of the CD of an active chiral medium was greatly enhanced as well and we could observe an important chiral mode splitting. We believe this research will accelerate the practical applications of superchiral field and the identification of chiral species.

**Acknowledgements:**

This research is financially supported by National Key Research and Development Project 2021YFA1400802, National Natural Science Foundation of China under the grant Nos. 12025402, 62125501, 11934012, 11974092, 61975041, Shenzhen Fundamental Research Projects JCYJ20210324120402006, JCYJ20200109112805990, JCYJ20200109113003946 and Fundamental Research Funds for Central Universities.